# Mn local moments prevent superconductivity in iron-pnictides Ba(Fe$_{1-x}$Mn$_x$)$_2$As$_2$

Y. Texier[1], Y. Laplace[1], P. Mendels[1,2], J. T. Park[3], G. Friemel[3], D. L. Sun[3], D. S. Inosov[3], C. T. Lin[3], J. Bobroff[1(a)]

[1] *Laboratoire de Physique des Solides, Univ. Paris-Sud, UMR8502, CNRS, F-91405 Orsay Cedex, France, EU*
[2] *Institut Universitaire de France, 103 Boulevard Saint-Michel, F-75005 Paris, France, EU*
[3] *Max Planck Institute for Solid State Research, Heisenbergstrase 1, D-70569 Stuttgart, Germany, EU*

12 June 2012

**Abstract** – $^{75}$As nuclear magnetic resonance (NMR) experiments were performed on Ba(Fe$_{1-x}$Mn$_x$)$_2$As$_2$ ($x_{Mn}$ = 2.5%, 5% and 12%) single crystals. The Fe layer magnetic susceptibility far from Mn atoms is probed by the $^{75}$As NMR line shift and is found similar to that of BaFe$_2$As$_2$, implying that Mn does not induce charge doping. A satellite line associated with the Mn nearest neighbours (n.n.) of $^{75}$As displays a Curie-Weiss shift which demonstrates that Mn carries a local magnetic moment. This is confirmed by the main line broadening typical of a RKKY-like Mn-induced staggered spin polarization. The Mn moment is due to the localization of the additional Mn hole. These findings explain why Mn does not induce superconductivity in the pnictides contrary to other dopants such as Co, Ni, Ru or K.

**Introduction.** – High temperature superconductivity can be induced by many different ways in iron-pnictides [1]. Starting from a semi-metal antiferromagnet such as the archetypal BaFe$_2$As$_2$, superconductivity can be achieved by hole doping (substituting K for Ba) [2], electron doping (substituting Co [3] or Ni [4] for Fe), applied pressure [5], or even by local disorder (substituting Ru for Fe) [6][7]. The resulting phase diagram is very similar in all these cases. It looks as if the only requirement to obtain superconductivity is to destroy sufficiently the antiferromagnetic (AF) state, whatever the mechanism at play.

Remarkably, there is an exception to this generic behavior: the substitution of Fe by hole-dopant atoms, namely Cr or Mn. Indeed, Mn and Cr substituted at Fe site do not lead to a superconducting state [8-11]. Neutron diffraction and magnetotransport measurements reveal that at low concentration, Mn or Cr substitutions reduce the ordering temperature $T_N$ by 6 K/% and the ordered moment amplitude remains unaffected [9][10]. Above $x$ = 30% for Cr and $x$ = 12% for Mn, new types of AF states appear whose $T_N$ even increases with further substitution. It is a G-type AF ordering for Cr doping [9], whereas Mn doping invokes an unusual tetragonal AF state with no orthorhombic distortion as in the parent compound [10]. Similar qualitative features are also observed in SrFe$_2$As$_2$ with Mn doping [11]. This asymmetry between the effects of hole (Mn, Cr) and electron (Co, Ni) dopants is surprising and still unexplained. The question arises as to why the phase diagram obtained under Mn or Cr hole doping is so different from that with out-of-plane K or Rb hole doping.

In this NMR study, we address this issue by showing that in the case of Mn, the additional holes do not delocalize, but act as local moments on Mn sites. These local moments are unable to suppress the AF state, preventing the appearance of superconductivity.

**Experimental details.** – We studied three Ba(Fe$_{1-x}$Mn$_x$)$_2$As$_2$ ($x_{Mn}$ = 2.5%, 5% and 12%) single crystals grown from self-flux in zirconia crucibles sealed in quartz ampoules under argon atmosphere, as described in [12]. $^{75}$As NMR was performed with applied magnetic field either at $H_0$ = 7.5 T or 13.5 T. All the spectra were obtained by standard Fourier Transform recombination after a $\pi/2 - \tau - \pi$ pulse sequence. The quality and homogeneity of the crystals is evidenced by the narrowness of the high temperature NMR spectrum. Indeed, previous studies showed that the $^{75}$As NMR line width in the BaFe$_2$As$_2$ family is enhanced by any small structural or electronic disorder [13]. Here, the line width at room temperature is as small as that observed in very homogeneous Co doped materials and much narrower than for Ru or K doping [7][14].

The NMR $^{75}$As nuclear spin $I$ = 3/2 is sensitive both to the local magnetic field and to the electric field gradient (EFG). When the crystal is oriented along a given crystallographic axis with respect to the applied magnetic field, the NMR spectrum splits into a central transition line and quadrupolar satellites. In this study, we focus only on the central line corresponding to the $1/2 \leftrightarrow -1/2$ nuclear transitions, whose position follows from

$$\nu = \frac{\gamma}{2\pi} H_0 + \frac{\gamma}{2\pi} H_0 \, {}^{75}K + \alpha \frac{\nu_Q^2}{\gamma H_0} \quad (1)$$

where $\gamma$ is the $^{75}$As gyromagnetic factor. The second term is proportional to the shift $^{75}K$ induced by hyperfine interactions with the electrons orbital and spin susceptibilities. The last term is the EFG contribution, where $\nu_Q$ is proportional to the main-axis electric field gradient, and $\alpha$ depends only on the orientation of the crystal with respect to $H_0$ and on the asymmetry of the EFG tensor. All measurements were performed after in-situ orientation of the crystal with field parallel to the crystallographic $c$ axis. For such an orientation, the EFG term cancels, since $\alpha = 0$ for symmetry reasons in the specific case of $^{75}$As in BaFe$_2$As$_2$. The resulting NMR spectrum is then expected to shift proportionally to $H_0$. This is indeed the case as shown for example in fig. 1 where the spectrum position and line shape are found to scale linearly with the applied field. Note the presence of a satellite, whose shift is also paramagnetic.

[a]E-mail: julien.bobroff@u-psud.fr

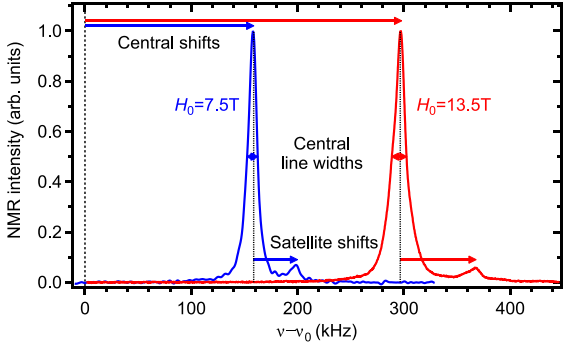

Fig. 1: (color online) Typical $^{75}$As NMR spectra obtained for $x_{Mn}$ = 2.5% for two different fields with $H_0 // c$ at $T$ = 200 K. $\nu_0 = \frac{\gamma}{2\pi} H_0$ is the reference frequency. All the spectral features (shift, linewidth, satellite) are found to be proportional to $H_0$.

The temperature dependences of the spectra for various Mn contents are displayed in fig. 2. The spectrum consists of a main line, which broadens with Mn content and with decreasing temperature, and a satellite (arrows in fig. 2), well defined for $x_{Mn}$ = 2.5% or 5%. For $x_{Mn}$ = 12%, the satellite merges with the central line and remains visible only at high temperatures. Spectra could not be measured below T=100K, since the spectral intensity drops to zero at $T$ = 110(5) K for $x_{Mn}$ = 2.5% and $T$ = 105(5) K for $x_{Mn}$ = 5%. This is a direct signature of a transition to the antiferromagnetic (AF) state: Fe moments freeze and strongly shift the NMR line out of our limited observation window. For $x$ = 12%, the intensity drop starts already at $T$ = 200 K, i.e. above the ordering temperature of the parent compound, and is progressive down to $T$ = 100 K, where the entire signal is lost, signaling a broader AF transition. The $x$ = 12% data will therefore not be taken into account in the following quantitative analysis in this study.

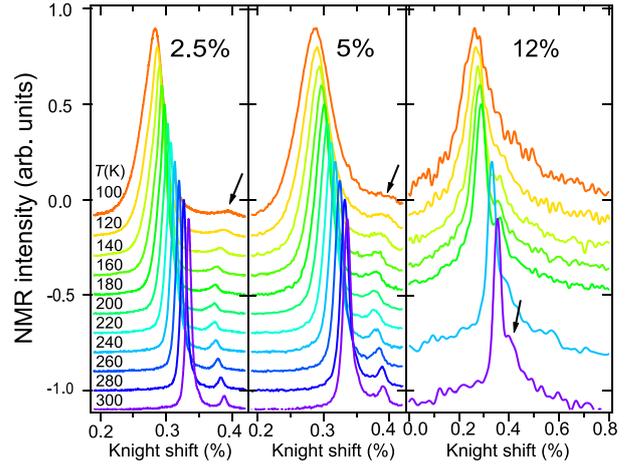

Fig. 2: (color online) $^{75}$As NMR spectra for $H_0 // c$ axis above $T_N$ for the $x_{Mn}$ = 2.5%, 5% and 12%. Curves are arbitrarily shifted vertically. Arrows mark the Mn n.n. satellite.

The observed modifications of the spectrum are characteristic of the presence of a local moment on Mn. Indeed, as demonstrated in the case of metals or correlated materials, impurity local moments induce a staggered spin polarization in their vicinity, which in turn broadens the spectrum and may even induce the appearance of satellites in the wings of the central line [15]. We will first discuss the central line position which monitors the intrinsic magnetic susceptibility far from Mn. We will then focus on the satellite which monitors the Mn immediate vicinity. Finally, we will use the main line broadenings to study the Mn induced spin polarization.

**Absence of hole doping** – We first focus on the regions far from Mn sites to address the average effect of Mn on the doping level of the Fe layer. These regions can be probed by the NMR shift $^{75}K$ of the main line. Indeed, since each $^{75}$As nucleus is coupled to its 4 n.n. Fe sites, $^{75}K$ is related to the local electron spin susceptibility of Fe layers $\chi_{Fe}$ through:

$$^{75}K_{central} = 4 A_{Fe}\chi_{Fe} + K_{orb} \quad (2)$$

where $A_{Fe}$ is the hyperfine coupling between one Fe and the $^{75}$As nucleus and $K_{orb}$ is a temperature-independent shift due to orbital effects. In fig. 3, $^{75}K$ is plotted versus temperature for the compounds of this study together with the parent BaFe$_2$As$_2$ [16], an electron doped Ba(Fe$_{0.895}$Co$_{0.105}$)$_2$As$_2$ [17] and a hole-doped Ba$_{0.68}$K$_{0.32}$Fe$_2$As$_2$ [18].

$^{75}K$ displays a smooth decrease with temperature typical of pnictides. It was found for various other materials that hole or electron doping shift the curve by an offset proportional to the doping level, positive for holes and negative for electrons [17,18]. So if every Mn$^{2+}$ atom was to dope the system with a hole as naively expected from its valence, this would lead to a positive offset of $K(T)$ increasing with Mn content. For example, in this hypothesis, the $x_{Mn}$ = 12% shift $^{75}K$ (blue in fig. 3)

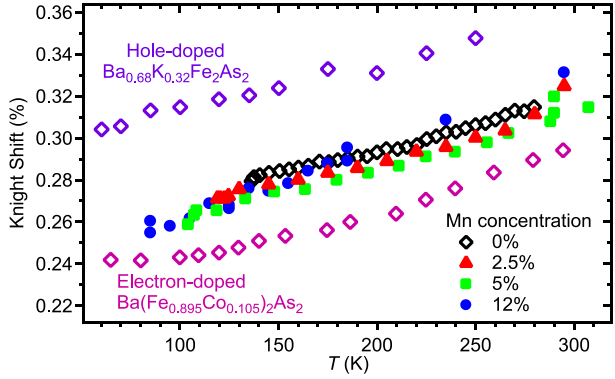

Fig. 3: (color online) The $^{75}$As Knight shift $^{75}K$ for the central line of Ba(Fe$_{1-x}$Mn$_x$)$_2$As$_2$ for $x_{Mn}$ = 2.5%, 5%, 12% (respectively triangles, squares and circles) is compared to that of the parent compound BaFe$_2$As$_2$ from [16] (black diamonds), the electron doped Ba(Fe$_{0.895}$Co$_{0.105}$)$_2$As$_2$ from [17] (pink diamonds), and the hole-doped Ba$_{0.68}$K$_{0.32}$Fe$_2$As$_2$ from [18] (purple diamonds).

should approach that of Ba$_{0.68}$K$_{0.32}$Fe$_2$As$_2$ (purple). Instead, the shifts for all Mn-substituted samples do not differ from the parent compound outside of the experimental error. This evidences that Mn substitution at the Fe site does not dope the system with either hole or electron, contrary to the cases of Co or K substitution on the Fe or Ba sites, respectively.

**Mn atoms carry a local moment.** – We now focus on the effect of Mn in its vicinity, which is reflected in the spectrum broadening and the appearance of a satellite. This satellite, marked by the arrows in fig. 2, is found to be independent of temperature. Its intensity increases with Mn content as shown in fig. 4, so that it is natural to associate this satellite with regions close to the Mn atoms. The intensity ratio $I$(satellite)/$I$(total) is a direct measure of the fraction of $^{75}$As atoms involved in the spatial regions associated to the satellite. This experimental ratio is almost equal to that expected for As sites with one n.n. Mn atom, as seen in fig. 4, where we also plot for comparison the average probabilistic number of As sites with exactly one or two Mn n.n., given by $4x(1-x)^3$ and $6x^2(1-x)^2$, respectively.

We thus conclude that the satellite corresponds to $^{75}$As nuclei whose 4 hyperfine coupled sites are occupied by 3 Fe and 1 Mn atoms. The deviation to the probabilistic law at large Mn content is not significant both because of the large uncertainties in the experimental determination of the satellite intensity, and because at such large Mn concentrations, segregation between rich and poor Mn regions could be expected, which would not obey the simple probabilistic model. We were not able to detect the satellite coming from $^{77}$As with two Mn n.n. most likely because of its very small intensity, typically one order of magnitude smaller than the one we observe.

The Mn n.n. satellite is caused by a modification of the $^{75}$As NMR resonance in the vicinity of Mn. From equation (1), it may come from a modification of $^{75}K$ or the local EFG. A similar satellite was observed in the case of Co doping be-cause the EFG was strongly modified [16]. But here, since the observed satellite frequency shift is found to be proportional to $H_0$ (fig.1), it must come from a modification of the shift instead. This modification comes from a local change in the electronic susceptibility due to Mn. We can estimate this change using the fact that each $^{75}$As nucleus participating in this satellite is hyperfine coupled to 1 adjacent Mn and 3 adjacent Fe leading to:

$$^{75}K_{sat} = 3A_{Fe} \cdot (\chi_{Fe} + \Delta\chi) + A_{Mn} \cdot \chi_{Mn} + K'_{orb} \quad (3)$$

where $\chi_{Fe}$ is the Fe susceptibility far from Mn, and $\Delta\chi$ accounts for a possible modification of this susceptibility near Mn. In a first rough analysis, we assume that this induced modification follows $\Delta\chi \propto \chi_{Mn}$ (see footnote [1]). Then using equation (2), we deduce the temperature dependence of the Mn local susceptibility:

$$\chi_{Mn}(T) \propto {}^{75}K_{sat}(T) - \frac{3}{4}{}^{75}K_{central}(T) + K''_{orb} \quad (4)$$

where $K''_{orb} = \frac{3}{4}K_{orb} - K'_{orb}$ is a $T$-independent orbital term. The resulting $T$-dependence of $\chi_{Mn}(T)$ is plotted in fig. 5 for $x_{Mn}$ = 2.5% and 5%. For $x$ = 12%, only high temperatures are displayed, because the satellite is not resolved at lower temperatures. $\chi_{Mn}(T)$ is found the same for $x_{Mn}$ = 2.5% and 5%, which justifies our analysis in terms of a local Mn susceptibility. The $T$-dependence of $\chi_{Mn}$ contrasts with that of the Fe far from Mn plotted in fig. 3, as the former increases while cooling instead of decreasing. $\chi_{Mn}(T)$ can be fitted by a Curie-Weiss law in $1/(T+\theta)$ (fig. 5) typical of a paramagnetic local moment. This Curie-Weiss behavior demonstrates the existence of a local moment on each Mn atom. The restricted temperature range accessible here ($T > 100$ K) prevents a precise determination of $\theta$ and adjustments within $-50$ K $< \theta < 50$ K are compatible with our data. The magnetic transition prevents measurements at lower temperatures, which could have allowed discriminating between a simple Curie law and a Curie-Weiss law with ferromagnetic or AF correlations. Similar local moments were also observed on Mn atoms in pure BaMn$_2$As$_2$ [20].

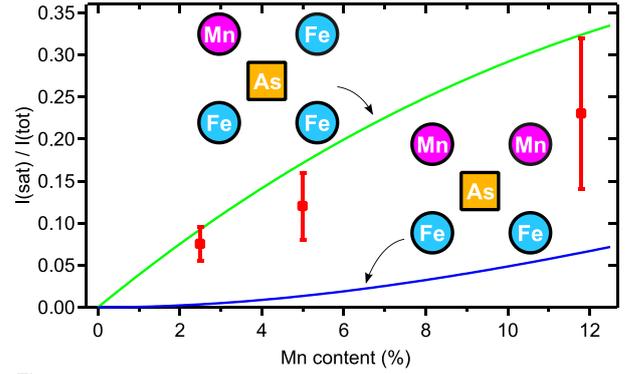

Fig. 4: (color online) Intensity ratio between the n.n. satellite and the whole $^{75}$As NMR spectrum as a function of the Mn content $x$. Green and blue lines are the expected ratios if the satellite corresponds to $^{75}$As with one or two n.n. Mn atoms, respectively.

---

[1] This hypothesis can be wrong if $\chi'(r)$ depends on temperature because of AF correlations [19].

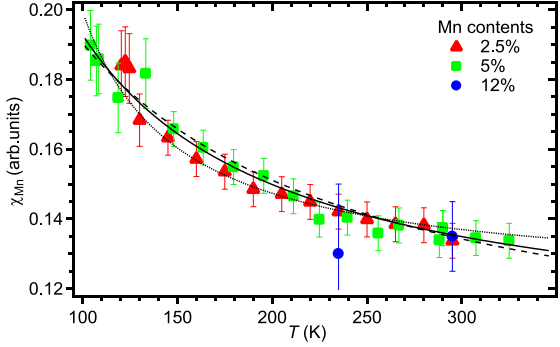

Fig. 5: (color online) *T*-dependence of the local Mn magnetic susceptibility $\chi_{\mathbf{Mn}}$ for various Mn contents evaluated using Eq. 4 as described in the text. Continuous, dashed and dotted lines are fits with a Curie-Weiss model: $\mathbf{y_0 + \frac{c}{T+\theta}}$ with $\boldsymbol{\theta}$ respectively fixed to 0 K, 50 K and −50 K.

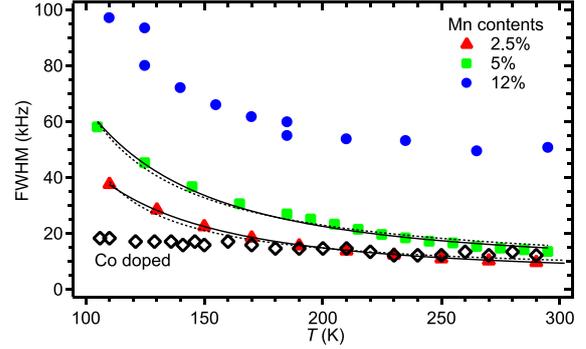

Fig. 6: (color online) Temperature dependence of the line width measured at half maximum height of the central line for Mn doping $x_{\mathrm{Mn}} = 2.5\%$, 5%, 12% and Co doping ($x_{\mathrm{Co}} = 6\%$) with $H_0 = 13.5$ T. Dashed line: fit with $\mathbf{y_0 + \frac{c}{T+\theta}}$. Continuous line: fit with $\mathbf{y_0 + \frac{c}{T^2}}$.

The EFG near Mn can also be determined, but using measurements where the field $H_0$ is not applied along the *c* axis so that the EFG does not vanish. For $x_{\mathrm{Mn}} = 2.5\%$, we measured the dependence of the central line and satellite positions versus field orientation. We could fit the satellite and main line shifts with a quadrupolar parameter $\nu_Q$ of 2030(80) kHz and 2490(50) kHz respectively. The EFG near Mn is decreased by about 20 % as compared to the pure material. This modification is rather small in comparison with the effect of other atomic substituents, such as Co, whose n.n. EFG is larger by more than a factor of 2 (5700 kHz) in Ba(Fe$_{1-x}$Co$_x$)$_2$As$_2$ [16], or Ru, which displays similar large EFG modifications [7]. We conclude that Mn modifies its EFG, i.e. its structural environment, only slightly, as compared to Co or Ru.

**Mn induced RKKY-like effects.** – Much like in metals or correlated systems, the Mn local paramagnetic moments induce a staggered spin polarization within the Fe layer. This is evidenced by the NMR central line width behavior: Mn impurities induce a characteristic additional symmetric broadening, which is reported in fig .6. The *T*-dependent part of the width $\Delta\nu$ is proportional to the Mn content $x_{\mathrm{Mn}}$ up to 5% and mimics the *T*-dependence of the Mn local susceptibility of fig. 5. This clearly contrasts with the *T*-independent behavior of a typical Co doped material plotted for comparison, which demonstrates here again that Mn carries a local moment while Co does not. The $x_{\mathrm{Mn}} = 12\%$ data should be taken with caution as already stressed.

The origin of this broadening lies in the response of the Fe spins to the Mn moment $g\mu_B\langle S_z\rangle$. This moment acts as a local field $\mathbf{H}(\mathbf{r}) \propto \langle S_z\rangle\delta(\mathbf{r})$ and induces at a distance $\mathbf{r}$ from the impurity an in-plane Fe spin polarization $s(\mathbf{r}) \propto \chi'(\mathbf{r})\langle S_z\rangle$ involving the spatial Fe susceptibility $\chi'(\mathbf{r})$. The NMR spectrum represents a histogram of the induced Fe-spin polarization as probed by the $^{75}$As sites, which is manifested in a symmetric broadening proportional to $\langle S_z\rangle x$ for small $x$. The broadening depends also on the shape and *T*-dependence of $\chi'(\mathbf{r})$. For example, in a metal, RKKY interactions lead to a Lorentzian broadening whose *T*-dependence follows that of the local moment, i.e. $1/(T+\Theta)$ [15]. In strongly correlated materials, impurities also induce a broadening, but with a different temperature dependence. In that case, $\chi'(\mathbf{r})$ reflects the magnetic correlations between spins, which may depend on temperature and combine with the Curie-Weiss behavior of the moment. In cuprates, it results in a typical $1/T^2$ behavior [19]. So the NMR response to impurities can be used to probe $\chi'(\mathbf{r})$ and decide for the importance of magnetic correlations.

However, in our study, the limited temperature range prevents from discriminating between a $1/(T+\theta)$ or a $1/T^2$ dependences of the line width (see fits in fig. 6). In the metallic Curie-Weiss scenario, we find a negative value for $\theta$ ($-50$ K $< \theta < -70$ K), synonymous of ferromagnetic correlations between Mn.

We also tried to discriminate between various models for $\chi'(r)$ by analyzing the NMR line shape. We performed simulations of the spectrum using a RKKY model for a 2D-metal and an AF correlated model like used for cuprates [15]:

$$\chi'_{\mathrm{RKKY}}(x,y) = \frac{A}{(x^2+y^2)} \cos\left(\frac{2\pi(x^2+y^2)^{1/2}}{\lambda}\right)$$

$$\chi'_{\mathrm{AF}}(x,y) = A(-1)^{x+y} e^{-\frac{(x^2+y^2)}{\lambda^2}}$$

where $\lambda$ is either inversely proportional to the Fermi wave vector $k_F$ in the former or to the AF correlation length in the latter, and *x* and *y* are in cell units. One-dimensional projections of these polarizations are plotted in the inset of fig. 7. We fixed the same *A* and $\lambda$ for all Mn contents. Each simulated spectrum is convoluted with a fixed Gaussian line to account for other sources of disorder also present in the parent compound. Typical simulations are compared to the experiment in fig. 7. Both models for $\chi'(r)$ reproduce well the central line broadening and the n.n. satellite for $x_{\mathrm{Mn}} = 2.5\%$ and 5%. For $x_{\mathrm{Mn}} = 12\%$, the simulations do not succeed to fully reproduce the experimental spectrum. But at such a large Mn concentration, strong overlap between the various Mn-induced effects must occur whose description goes beyond the scope of this study.

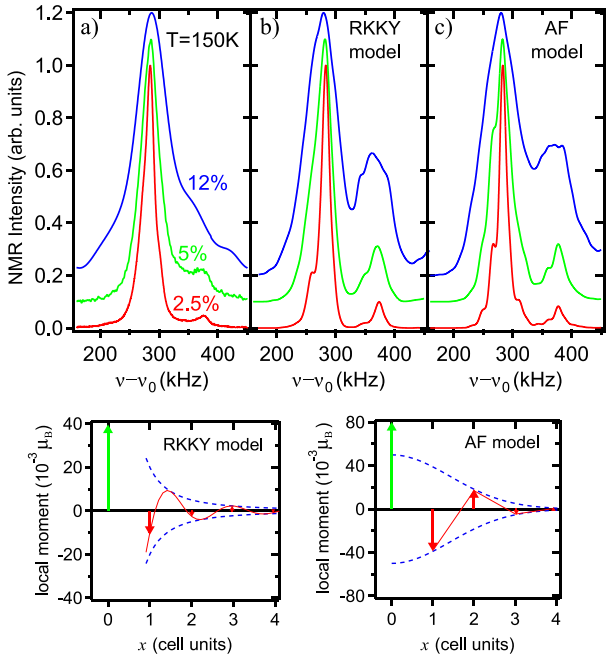

Fig. 7: (color online) a) NMR spectrum for $x_{Mn}$ = 2.5%, 5%, 12% at $T$ = 150 K. b), c) Simulations of the spectrum using respectively RKKY and AF polarization models d), e) One-dimensional projection of the spin polarizations (red) induced by the Mn local moment (green). See text for the model of the envelope (blue line).

The spectral shapes and temperature dependences are thus well explained by the existence of Mn local moments, which induce an alternating spin polarization in the Fe layer. But we are not able to discriminate between various shapes of that polarization, i.e. of $\chi'(r)$.

**Discussion and conclusion.** – From our experiment, Mn$^{2+}$ substituted for Fe$^{2+}$ does not modify the doping of the BaFe$_2$As$_2$ pnictides and maintains their semi-metallic nature with equal concentration of electron- and hole-type charge carriers. The additional Mn hole remains localized and results in the formation of a local magnetic moment, on the contrary to Co or Ni doping. This moment in turn couples to the Fe electronic band and induces an alternating spin polarization; a common behavior encountered both in weakly or strongly correlated metals.

At small concentration (a few %), Mn neither changes the electronic doping nor induces a strong structural or electronic disorder from our Knight-shift and EFG analysis. The Mn local moments are not expected to modify the multi-band Fermi surface strongly, nor to severely destroy the Neel ordering. This explains why the phase diagram of Ba(Fe$_{1-x}$Mn$_x$)$_2$As$_2$ is so different from that of Co, Ni, K or Ru doped pnictides.

At $x_{Mn} \gtrsim 10\%$, the change into another type of magnetic ordering with similar Neel temperatures probably stems from an intricate situation involving Fe and Mn moments in similar proportions and possibly represents a percolation-type crossover to a state, where direct interactions between Mn moments are no longer negligible. Actually, the Mn phase diagram merely describes the evolution of an AF semi-metal in the presence of local moments. Furthermore, even if the Neel state was to be destroyed, superconductivity would probably be strongly weakened or even completely destroyed because the Mn local moments would then act as magnetic pair-breakers of the superconducting Cooper pairs.

It is not clear why Mn and Cr located on the hole-doping side of Fe form a local moment while Ni or Co on the electron-doping side do not. It is probably due to an interplay between the multiband complex structure of the material, the strong Hund coupling, and the impurity level position, which makes the additional Mn hole form a local moment, on the contrary to Co or Ni. Band structure calculations are needed to explain this more quantitatively.

Finally, our study opens a new route towards measurements of the electronic correlations in these materials through $\chi'(r)$ using the Mn induced NMR broadenings. Mn local impurities could in fact be used to reveal the nature and importance of the correlations of these compounds, as has been already done in various other unconventional materials.

∗ ∗ ∗

We thank H. ALLOUL, V. BROUET, F. BERT, A. BORIS, D. EFREMOV and A. YARESKO for fruitful and stimulating discussions and M. LE TACON for initiating this collaboration. This work has been supported by the ANR Pnictides and by the DFG within the Schwerpunktprogramm 1458, under Grand N°. BO3537/1-1.